\documentclass[conference]{IEEEtran}
\IEEEoverridecommandlockouts
\usepackage{algorithm,algorithmic}

\makeatletter
\usepackage[letterpaper, left=1in, right=1in, bottom=0.9in, top=0.75in]{geometry}

\newcommand\fs@betterruled{%
  \def\@fs@cfont{\bfseries}\let\@fs@capt\floatc@ruled
  \def\@fs@pre{\vspace*{5pt}\hrule height.8pt depth0pt \kern2pt}%
  \def\@fs@post{\kern2pt\hrule\relax}%
  \def\@fs@mid{\kern2pt\hrule\kern2pt}%
  \let\@fs@iftopcapt\iftrue}
\floatstyle{betterruled}
\restylefloat{algorithm}
\makeatother

\newcommand{\squeezeup}{\vspace{-1mm}}

\usepackage{cite}
\usepackage{amsmath,amssymb,amsfonts}
\usepackage{algorithmic}
\usepackage{graphicx}
\usepackage{textcomp}
\DeclareMathOperator*{\argmin}{arg\,min}
    
\begin{document}

\title{A Unified Beamforming and A/D Self-Interference Cancellation Design for Full Duplex MIMO Radios}

\author{\IEEEauthorblockN{Md Atiqul Islam\IEEEauthorrefmark{2}, George C. Alexandropoulos\IEEEauthorrefmark{4}, and Besma Smida\IEEEauthorrefmark{2}}
\IEEEauthorblockA{{\IEEEauthorrefmark{2}Department of Electrical and Computer Engineering, University of Illinois at Chicago, USA}\\
\IEEEauthorrefmark{4}Department of Informatics and Telecommunications, National and Kapodistrian University of Athens, Greece\\
emails: \{mislam23,smida\}@uic.edu, alexandg@di.uoa.gr
}}
\maketitle

\begin{abstract}
In this paper, we focus on reduced complexity full duplex Multiple-Input Multiple-Output (MIMO) systems and present a joint design of digital transmit and receive beamforming with Analog and Digital (A/D) self-interference cancellation. We capitalize on a recently proposed multi-tap analog canceller architecture, whose number of taps does not scale with the number of transceiver antennas, and consider practical transmitter impairments for the full duplex operation. Particularly, transmitter IQ imbalance and nonlinear power amplification are assumed via relevant realistic models. Aiming at suppressing the residual linear and nonlinear self-interference signal below the noise floor, we propose a novel digital self-interference cancellation technique that is jointly designed with the configuration of the analog taps and digital beamformers. Differently from the state of the art, we design pilot-assisted estimation of all involved wireless channels. Our representative Monte Carlo simulation results demonstrate that our unified full duplex MIMO design exhibits higher self-interference cancellation capability with less analog taps compared to available techniques, which results in improved achievable rate and bit error performance.
\end{abstract}


\section{Introduction}
Recent advances in Full Duplex (FD) communication technology prove the potential of a twofold increase in spectral efficiency over the conventional frequency- and time-division duplex systems through simultaneous uplink and downlink communication in the same frequency resources \cite{sabharwal2014band,bharadia2013full,smida2017reflectfx,duarte2012experiment}. Exploitation of Multiple-Input Multiple-Output (MIMO) communication provides an efficient system performance boost due to increasing the system's spatial degrees of freedom (DoF) offered by the plurality of Transmitter (TX) and Receiver (RX) antennas \cite{riihonen2011mitigation,everett2016softnull,alexandropoulos2017joint,bharadia2014full,masmoudi2017channel,anttila2014modeling}. Thus, enabling FD in conjunction with MIMO operation can achieve the demanding throughput requirements of fifth Generation (5G), and beyond, wireless communication systems with limited spectrum resources.\par

The simultaneous transmission and reception in FD systems results in practice in an in-band Self-Interference (SI) signal at the RX side due to the limited isolation between the TX and RX blocks. Since the propagation path for this signal is shorter than any other signal-of-interest from a far-end communicating node, the SI signal at RX is many times stronger \cite{sabharwal2014band}. Therefore, SI cancellation techniques are required to suppress SI below the RX noise floor. In FD MIMO systems, the suppression techniques are particularly challenging due to higher SI components as a consequence of the increased number of antennas. Conventional SI suppression techniques in Single-Input Single-Output (SISO) systems combine propagation domain isolation, analog domain suppression, and digital cancellation\cite{bharadia2013full,korpi2014widely}. Propagation domain isolation is accomplished by pathloss or antenna directionality\cite{korpi2017compact}, whereas analog domain suppression is achieved by subtracting a processed copy of the TX signal from the RX inputs to avoid saturation of the RX Radio Frequency (RF) chain\cite{sabharwal2014band,khaledian2018inherent}. Analog SI cancellation in FD MIMO systems can be implemented through SISO replication. However, the hardware requirements for such an approach scale with the number of TX/RX antennas, rendering the implementation of analog SI a core design bottleneck\cite{george2018journal}. The authors in \cite{riihonen2011mitigation,everett2016softnull} presented spatial suppression techniques that alleviate the need for analog SI cancellation, which was replaced solely by digital TX/RX beamforming. In \cite{alexandropoulos2017joint}, a joint design of multi-tap analog cancellation and TX/RX beamforming, where the number of taps does not scale with the product of TX and RX antenna elements, was proposed. This approach reduces the hardware complexity of conventional FD MIMO designs \cite{bharadia2014full,masmoudi2017channel,anttila2014modeling} based on analog SI cancellation  through SISO replication.

In practical FD MIMO systems, there exist impairments contaminating the signal-of-interest when it is traversing through the TX block. These include IQ imbalances and Power Amplifier (PA) nonlinearities inserted in the transceiver RF chain. Such impairments need to be accurately estimated in order to be extracted from the residual signal after analog SI cancellation. Conventionally in FD SISO systems, the latter nonlinear components are suppressed in the digital domain. The effect of PA nonlinearities is analyzed to formulate digital cancellation techniques in \cite{bharadia2013full,ahmed2015all}, whereas the impact of IQ imbalances and resulting image components is investigated in \cite{korpi2014widely}. A comprehensive digital cancellation approach for single-antenna FD SISO systems considering both transmitter and receiver nonlinearities and IQ imbalances is presented in\cite{islam2019comprehensive}. However, in FD MIMO, their impact will be pronounced especially when the TX power is high \cite{bharadia2014full,anttila2014modeling}.

 In this paper, we present a joint design of digital TX/RX beamforming with Analog and Digital (A/D) SI cancellation for practical FD MIMO systems with TX IQ imbalances and PA nonlinearities. Building upon the reduced complexity multi-tap analog canceller architecture of \cite{alexandropoulos2017joint}, we design a novel digital SI cancellation technique that is jointly designed with the configuration of the analog taps and digital TX/RX beamformers. Selected simulation results showcase the interplay of the individual components of our proposed FD MIMO approach when TX impairments are considered. It is also shown that our design exhibits improved performance with less analog canceller taps compared to available approaches.\par
\textit{Notation:} Vectors and matrices are denoted by boldface lowercase and boldface capital letters, respectively. The transpose, Hermitian transpose, and conjugate of $\mathbf{A}$ are denoted by $\mathbf{A}^{\rm T}$, $\mathbf{A}^{\rm H}$, and $\mathbf{A}^*$, respectively, and $\det(\mathbf{A})$ is $\mathbf{A}$'s determinant, while $\mathbf{I}_{n}$ ($n\geq2$) is the $n\times n$ identity matrix and $\mathbf{0}_{m\times n}$ ($m\geq2$ and $n\geq1$) represents the $m\times n$ matrix with all zeros. $\|\mathbf{a}\|$ stands for the Euclidean norm of $\mathbf{a}$, $\mathbf{a}^{\circ n}$ denotes the Hadamard power operation to a factor of n, operand $\odot$ represents the Hadamard entry-wise product, ${\rm col}\{\mathbf{a}_1,\mathbf{a}_2,\ldots,\mathbf{a}_n\}$ denotes a column vector after vertically concatenating vectors $\mathbf{a}_1,\mathbf{a}_2,\ldots,\mathbf{a}_n$, and ${\rm diag}\{\mathbf{a}\}$ denotes a square diagonal matrix with $\mathbf{a}$'s elements in its main diagonal. $[\mathbf{A}]_{i,j}$, $[\mathbf{A}]_{(i,:)}$, and $[\mathbf{A}]_{(:,j)}$ represent $\mathbf{A}$'s $(i,j)$-th element, $i$-th row, and $j$-th column, respectively, while $[\mathbf{a}]_{i}$ denotes the $i$-th element of $\mathbf{a}$. $\mathbb{C}$ represents the complex number set, $\mathbb{E}\{\cdot\}$ is the expectation operator, and $|\cdot|$ denotes the amplitude of a complex number.

\section{System and Signal Models}
\begin{figure*}[!t]
\centering
\includegraphics[width=\textwidth]{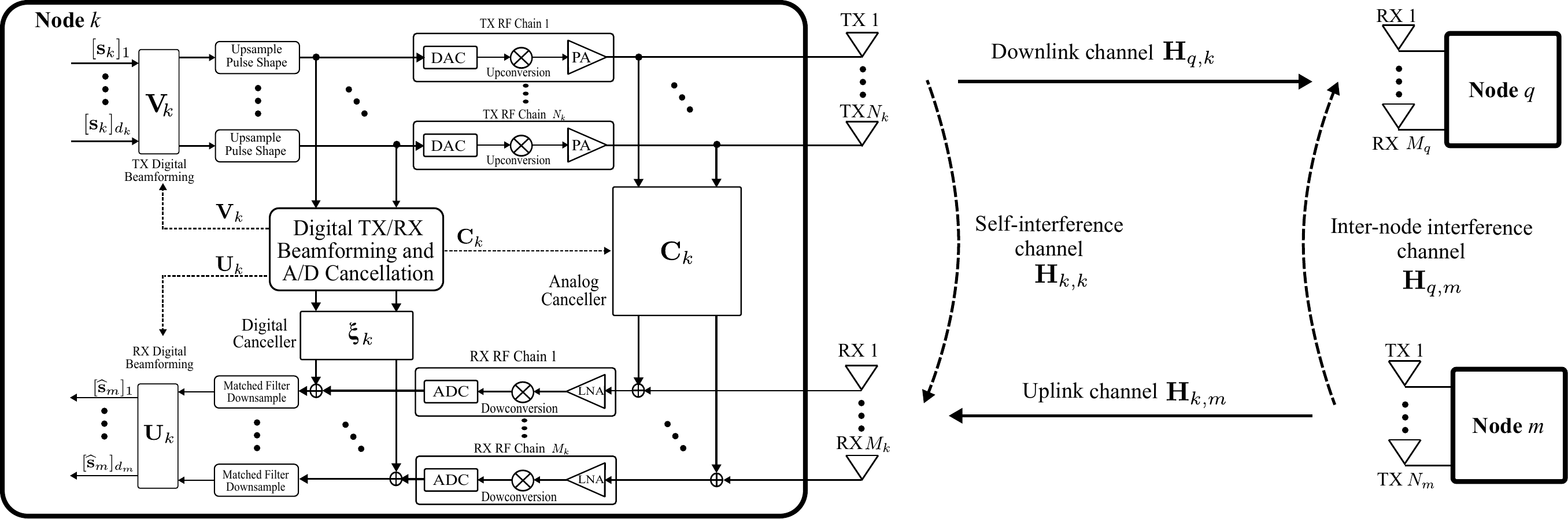}
\caption{The considered three-node communication system and the proposed FD MIMO architecture. The FD MIMO node $k$ incorporates processing blocks for analog SI cancellation and digital TX/RX beamforming similar to \cite{alexandropoulos2017joint}, as well as for digital cancellation of residual SI. All these blocks are jointly designed for efficient FD operation. The half-duplex multi-antenna nodes $q$ and $m$ communicate with node $k$ in the downlink and uplink directions, respectively.}
\label{fig: transceiver}
\squeezeup
\end{figure*}
The considered wireless communication system of Fig$.$~\ref{fig: transceiver} is comprised of a FD MIMO node $k$ wishing to communicate concurrently with two half duplex multi-antenna nodes: node $q$ in the DownLink (DL) and node $m$ in the UpLink (UL) communication. The FD MIMO node $k$ is assumed to be equipped with $N_k$ TX and $M_k$ RX antenna elements. Each TX antenna is attached to a dedicated RF chain; similarly holds for the RX RF chain and their attached antennas. The half duplex multi-antenna nodes $q$ and $m$ are assumed to have $M_q$ and $N_m$ antennas, respectively, with each of their antennas connected to a respective RF chain. All three nodes are considered capable of performing digital BF; for simplicity, we assume hereinafter that digital TX/RX beamforming at the focused FD MIMO node $k$ is realized with linear filters. In particular, we assume that node $k$ makes use of the precoding matrix $\mathbf{V}_k\in\mathbb{C}^{N_k\times d_k}$ for processing its symbol vector $\mathbf{s}_k\in\mathbb{C}^{d_k\times1}$ at baseband. The dimension of $\mathbf{s}_k$ satisfies $d_k\leq\min\{M_q,N_k\}$, which complies with the available spatial degrees of freedom for the downlink $M_q\times N_k$ MIMO channel. Similarly, node $m$ processes its symbol vector $\mathbf{s}_m\in\mathbb{C}^{d_m\times1}$ with a baseband precoding matrix $\mathbf{V}_m\in\mathbb{C}^{N_m\times d_m}$, where $d_m\leq\min\{M_k,N_m\}$. Without loss of generality, we assume that $\mathbf{V}_k$ and $\mathbf{V}_m$ have unit norm columns, and we use notations $\mathbf{x}_{k} \triangleq\mathbf{V}_k\mathbf{s}_k\in\mathbb{C}^{N_k\times1}$ and $\mathbf{x}_{m}\triangleq\mathbf{V}_{m}\mathbf{s}_{m}\in\mathbb{C}^{N_m\times1}$ for the output signals of the TX baseband blocks at nodes $k$ and $m$, respectively.

\subsection{TX Signal Model Incorporating Practical Impairments}\label{sec:Impairments}
As shown in node $k$ of Fig$.$~\ref{fig: transceiver}, the $N_k$ baseband signals in $\mathbf{x}_{k}$ are fed to the TX RF chains for upconversion and amplification. In this section, we introduce a baseband equivalent model for each of these chains incorporating IQ imbalances and PA nonlinearities. We assume that all TX RF chains are identical. For simplification, we do not consider such impairments for the TX of node $m$, which is instead assumed to be linear\footnote{In general practical TXs, nonlinear distortion components are $30$ dB lower than the linear signal\cite{bharadia2013full}. For typical UL pathloss of $110$ dB, these components fall well below the RX noise floor. Thus, ignoring these nonlinear components in the UL signal model will have a small impact on link performance.}.

After entering the TX RF chains of node $k$, the baseband signal goes through the IQ mixer for upconversion to the carrier frequency. As stated in \cite{korpi2014widely}, IQ imbalances in practical mixers induce a mirror image of the original signal with certain image attenuation.
By denoting $[\mathbf{x}_k]_i$ as one sample of the baseband signal at the input of the $i$-th TX RF chain of node $k$, the IQ mixer output can be written as \cite{korpi2014widely}
\begin{equation}\label{eq1}
    \begin{split}
        [\mathbf{x}_k]_i^{\rm IQ} \triangleq \mu_{1}[\mathbf{x}_k]_i + \mu_{2}[\mathbf{x}_k]_i^{*},
    \end{split}
\end{equation}
where $\mu_{1}\triangleq(1+g e^{-j\theta})/2$ and $\mu_{2}\triangleq(1-g e^{j\theta})/2$ with $g$ and $\theta$ representing the gain and phase imbalances, respectively. The Image Rejection Ratio, defined as ${\rm IRR}\triangleq\left|\mu_{1}/\mu_{2}\right|^2$, represents the strength of the IQ induced conjugate term\cite{korpi2014full}.

Before transmission, the upconverted signal is fed into the PA for amplification, while satisfying the TX power constraint. Note that practical PAs exhibit varying degrees of nonlinearity. For brevity, we consider a quasi memoryless PA model inducing only an odd-order nonlinearity, as the even-power harmonics lie out of band and will be cutoff by the RF low pass filter at RX \cite{korpi2014widely, zhou2005baseband}. For this PA model, the baseband equivalent of each of its output samples is given using \eqref{eq1} as

\begin{equation}\label{eq:PA_output}
    \begin{split}
        [\mathbf{x}_k]_i^{\rm PA} \!& \triangleq\!  g_{1,i}[\mathbf{x}_k]_i\! + g_{2,i}[\mathbf{x}_k]_i^*\! + g_{3,i}[\mathbf{x}_k]_i^3\!+\! g_{4,i}[\mathbf{x}_k]_i^2[\mathbf{x}_k]_i^*
        \\& \quad + g_{5,i}[\mathbf{x}_k]_i([\mathbf{x}_k]_i^*)^2 + g_{6,i}([\mathbf{x}_k]_i^*)^3, 
    \end{split}
\end{equation}
where the six gain components 
are derived as 
\begin{equation}   \label{eq:small_gs} 
    \begin{split}
        g_{1,i}&\triangleq\mu_{1}\nu_{1},\; g_{2,i} \triangleq \mu_{2}\nu_{1},\; g_{3,i}\triangleq \mu^2_{1}\mu^{*}_{2}\nu_{3},\\
        g_{4,i}&\triangleq (2|\mu_{1}|^2\mu_{1}\!+\!|\mu_{2}|^2\mu_{1})\nu_{3},\\
        g_{5,i}&\triangleq (2|\mu_{1}|^2\mu_{2}\!+\!|\mu_{2}|^2\mu_{2})\nu_{3},\,g_{6,i}\triangleq\mu^{*}_{1}\mu^{2}_{2}\nu_{3},
    \end{split}
\end{equation}
where $\nu_{1}$ denotes the PA linear gain, $\nu_{3}\triangleq{\nu_{1}}{/}{{(\rm IIP3)}^2}$ is the gain of the third order nonlinear distortions, and ${\rm IIP3}$ is the third order Input-referred Intercept Point of the PA\cite{gu2005rf}.
We have considered up to third order distortions, which represents the strongest nonlinearity at the PA output in practice\cite{korpi2014widely}.\par
Based on \eqref{eq:PA_output} and after some algebraic manipulations, the baseband representation of the impaired transmitted signal from the $N_k$ TX antennas of FD MIMO node $k$ at the DL direction can be expressed as
\begin{equation}\label{eq: signal_DL}
    \begin{split}
        \mathbf{\widetilde{x}}_k &\triangleq\mathbf{G}_{1,k}\mathbf{x}_k+\mathbf{G}_{2,k}\mathbf{x}_k^* + \mathbf{G}_{3,k}\mathbf{x}_k^{\circ 3} + \mathbf{G}_{4,k}\mathbf{x}_k^{\circ 2}\odot\mathbf{x}_k^* \\&\quad+ \mathbf{G}_{5,k}\mathbf{x}_k\odot(\mathbf{x}^*_k)^{\circ 2}  +\mathbf{G}_{6,k}(\mathbf{x}_k^{*})^{\circ 3}\\
        &= \mathbf{G}_{1,k}\mathbf{V}_k\mathbf{s}_k + \boldsymbol{\delta}_k=\mathbf{G}_k \boldsymbol{\psi}_k,
    \end{split}
\end{equation}
where $\mathbf{G}_{1,k}\triangleq\text{diag}\{g_{1,1},g_{1,2},\dots,g_{1,N_k}\}$ is the power allocation matrix of the linear components and $\mathbf{G}_{\ell,k}\triangleq\text{diag}\{g_{\ell,1},g_{\ell,2},\dots,g_{\ell,N_k}\}$ for $\ell=2,3,\ldots,6$ are the coefficient matrices for the remaining nonlinear components defined in similar way. In \eqref{eq: signal_DL}, we also introduce the notation $\boldsymbol{\delta}_k\in\mathbb{C}^{N_k\times 1}$ for the nonlinear and conjugate signal components, as well as the augmented power allocation matrix $\mathbf{G}_k\in\mathbb{C}^{N_k\times 6N_k}$, and the vertically arranged signal vector $\boldsymbol{\psi}_k\in\mathbb{C}^{6N_k\times 1}$ including the image and nonlinear components. The latter matrices are given by
\begin{equation}\label{eq:augment_signal}
    \begin{split}
        \mathbf{G}_k &\triangleq \left[\mathbf{G}_{1,k}\,\, \mathbf{G}_{2,k}\,\,\mathbf{G}_{3,k}\,\,\mathbf{G}_{4,k}\,\mathbf{G}_{5,k}\,\,\mathbf{G}_{6,k}\right],\\
        \boldsymbol{\psi}_k&\triangleq\text{col}\{\mathbf{x}_k,\mathbf{x}_k^*,\mathbf{x}_k^{\circ 3},\mathbf{x}_k^{\circ 2}\odot\mathbf{x}_k^*,\mathbf{x}_k\odot(\mathbf{{x}}^*_k)^{\circ 2},(\mathbf{{x}}_k^{*})^{\circ 3}\}.
    \end{split}
\end{equation}
The signal transmissions from nodes $k$ and $m$ are power limited to ${\rm P}_k$ and ${\rm P}_m$, respectively. Specifically, the DL signal is such that $\mathbb{E}\{\|\mathbf{G}_{1,k}\mathbf{V}_k\mathbf{s}_k +\boldsymbol{\delta}_k\|^2\}\leq {\rm P}_k$, whereas the UL signal is constrained as $\mathbb{E}\{\|\mathbf{G}_{1,m}\mathbf{V}_m\mathbf{s}_m\|^2\}\leq {\rm P}_m$ with $\mathbf{G}_{1,m}\in\mathbb{C}^{N_m\times N_m}$ representing the diagonal (linear) power allocation matrix of the precoded signal components. 

\subsection{Baseband Model for the RX Signals}
Following the definitions in Sec$.$~\ref{sec:Impairments}, the baseband received signal $\mathbf{y}_{q}\in\mathbb{C}^{M_q\times 1}$ at node $q$ is mathematically expressed as 
\begin{equation}\label{Eq:Received_q}
\mathbf{y}_{q} \triangleq \mathbf{H}_{q,k}\mathbf{G}_{1,k}\mathbf{V}_k\mathbf{s}_{k} + \mathbf{H}_{q,k}\boldsymbol{\delta}_k+ \mathbf{H}_{q,m}\mathbf{G}_{1,m}\mathbf{V}_{m}\mathbf{s}_{m} + \mathbf{n}_{q}, 
\end{equation}
where $\mathbf{H}_{q,k}\in\mathbb{C}^{M_q\times N_k}$ is the DL channel matrix, $\mathbf{H}_{q,m}\in\mathbb{C}^{M_q\times N_m}$ denotes the channel matrix for inter-node interference (i$.$e$.$, between nodes $q$ and $m$), and $\mathbf{n}_{q}\in\mathbb{c}^{M_q\times 1}$ represents the additive white Gaussian noise (AWGN) vector at node $q$ with covariance matrix $\sigma_{q}^{2}\mathbf{I}_{M_q}$. Assuming no inter-node interference between node $q$ and $m$ (i$.$e$.$, $\mathbf{H}_{q,m}=\mathbf{0}_{q,m}$ in \eqref{Eq:Received_q}) due to appropriate node scheduling\cite{alexandropoulos2016user,atzeni2016performance}, the linearly processed estimated symbol vector $\hat{\mathbf{s}}_k \in \mathbb{C}^{d_k\times 1}$ for $\mathbf{s}_k$ at the half duplex multi-antenna node $q$ is expressed as
\begin{equation}\label{Eq:est_s_q}
    \begin{split}
        \hat{\mathbf{s}}_k \triangleq  \mathbf{U}_q\left(\mathbf{H}_{q,k}\mathbf{G}_{1,k}\mathbf{V}_k\mathbf{s}_{k} + \mathbf{H}_{q,k}\boldsymbol{\delta}_k + \mathbf{n}_{q}\right),
    \end{split}
\end{equation}
where $\mathbf{U}_q\in \mathbb{C}^{M_q\times d_k}$ denotes the optimum combining matrix.

Upon signal reception at the FD MIMO node $k$, analog SI cancellation is first applied to the signals received at its RX antenna elements before these signals enter to the RX RF chains, as shown in Fig$.$~\ref{fig: transceiver}. Similar to \cite{alexandropoulos2017joint} we use notation $\mathbf{C}_k\in\mathbb{C}^{M_k\times N_k}$ to denote the signal processing realized by the multi-tap analog canceller. The baseband representation for the $M_k\times 1$ complex-valued output signal of the analog canceller at node $k$ can be expressed as
\begin{equation}\label{Eq:y_k_AC}
\widetilde{\mathbf{y}}_{k} \triangleq \mathbf{C}_k\left(\mathbf{G}_{1,k}\mathbf{V}_k\mathbf{s}_k+ \boldsymbol{\delta}_k\right).
\end{equation}
In order to suppress the residual SI after applying analog cancellation, we further consider digital SI cancellation from the received signals at the outputs of the RX RF chains. We note that for the practical case of channel estimation and TX impairments considered in this paper, digital SI cancellation is necessary, as will be shown in the simulation results. After A/D SI cancellation, the digitally converted and downsampled received signals are linearly processed in baseband by the combining matrix $\mathbf{U}_k\in\mathbb{C}^{d_m\times M_k}$, which we assume having unit norm columns. Then, the estimated symbol vector $\hat{\mathbf{s}}_m\in\mathbb{C}^{d_m\times 1}$ for $\mathbf{s}_m$ at the FD MIMO node $k$ is derived as
\begin{equation}\label{Eq:Estimated_m} 
\hat{\mathbf{s}}_m \triangleq  \mathbf{U}_k\left(\widetilde{{\mathbf{y}}}_k + \mathbf{y}_k+\overline{\mathbf{y}}_k +\boldsymbol{\xi}_{k}+\mathbf{n}_{k}\right), 
\end{equation}
where the complex-valued $M_k$-element vectors $\mathbf{y}_k$, $\overline{\mathbf{y}}_k$, and $\boldsymbol{\xi}_{k}$ are the baseband representations of the received signal of interest, received SI signal, and digital SI cancellation signal, respectively, at node $k$. In addition, $\mathbf{n}_{k}\in\mathbb{C}^{M_k\times 1}$ denotes the received AWGN vector at node $k$ with covariance matrix $\sigma_{k}^{2}\mathbf{I}_{M_k}$. Vector $\mathbf{y}_k$ including the signal of interest is given by  
\begin{equation}\label{Eq:y_k_SoI}
\mathbf{y}_{k} \triangleq \mathbf{H}_{k,m}\mathbf{G}_{1,m}\mathbf{V}_m\mathbf{s}_m,
\end{equation}
where $\mathbf{H}_{k,m}\in\mathbb{C}^{M_k\times N_m}$ is the UL channel matrix (i$.$e$.$, between nodes $k$ and $m$), while $\overline{\mathbf{y}}_k$ representing the received SI signal with the TX impairments is expressed as 
\begin{equation}\label{Eq:y_k_SI}
\overline{\mathbf{y}}_{k} \triangleq \mathbf{H}_{k,k}\left(\mathbf{G}_{1,k}\mathbf{V}_k\mathbf{s}_k+ \boldsymbol{\delta}_k\right)
\end{equation}
with $\mathbf{H}_{k,k}\in\mathbb{C}^{M_k\times N_k}$ being the SI wireless channel seen at the RX antennas of node $k$ due to its own DL transmission. Signal $\boldsymbol{\xi}_{k}\in\mathbb{C}^{M_k\times 1}$ for cancelling the residual SI signal in the digital domain will be described in the sequel.

\section{Proposed Unified FD MIMO Design}\label{sec: Proposed analog and digital}
In this section, we describe the joint design of digital TX/RX beamforming with A/D SI cancellation. We first extend the approach of \cite{alexandropoulos2017joint} to the case of pilot-estimated channels, which we then co-design with a novel digital cancellation technique to further suppress the residual SI signal.

\subsection{Digital TX/RX Beamforming and Analog SI Cancellation}\label{subsec: analog_canellation}
Suppose that the UL, DL, and SI wireless channels in the considered system of Fig$.$~\ref{fig: transceiver} are estimated using pilot signals as $\widehat{\mathbf{H}}_{k,m}$, $\widehat{\mathbf{H}}_{q,k}$, and $\widehat{\mathbf{H}}_{k,k}$, respectively. Using these estimations and the representations for the analog canceller and digital TX/RX beamformers, estimates for the achievable UL and DL rates can be respectively calculated as
\begin{align}\label{eq:rates_eq}
        \nonumber\widehat{\mathcal{R}}_{\text{UL}}\!&\triangleq\log_2\!\left(\!\det\!\left(\mathbf{I}_{M_k}\!+\! \|\mathbf{U}_k\widehat{\mathbf{H}}_{k,m}\mathbf{G}_{1,m}\mathbf{V}_m\|^2{\widehat{\mathbf{W}}_k}^{-1}\right)\!\right)\!,\\
        \widehat{\mathcal{R}}_{\text{DL}}\!&\triangleq\log_2\left(\det\left(\mathbf{I}_{M_q}\!+\! \|\mathbf{U}_q\widehat{\mathbf{H}}_{q,k}\mathbf{G}_{1,k}\mathbf{V}_k\|^2{\widehat{\mathbf{W}}_q}^{-1}\right)\right)\!,
\end{align}
where $\widehat{\mathbf{W}}_k$ and $\widehat{\mathbf{W}}_q$ denote the estimated covariances matrices of interference plus noise at multi-antenna nodes $k$ and $q$, respectively, which can be computed as
\begin{equation}\label{eq:covar_eq}
    \begin{split}
        \widehat{\mathbf{W}}_k \!&\triangleq\! \|\mathbf{U}_k(\widehat{\mathbf{H}}_{k,k}\!+\!\mathbf{C}_k)\!\left(\mathbf{G}_{1,k}\mathbf{V}_k + \boldsymbol{\delta}_k\right)\|^2 \!+\! \sigma_{k}^{2}\|\mathbf{U}_k\|^2,\\
        \widehat{\mathbf{W}}_q\! &\triangleq\!\|\mathbf{U}_q\widehat{\mathbf{H}}_{q,k}\boldsymbol{\delta}_k\|^2 \!+\!\sigma_{q}^{2}\|\mathbf{U}_q\|^2.
    \end{split}
\end{equation}
The latter expressions have been obtained from \eqref{Eq:est_s_q} and \eqref{Eq:Estimated_m} assuming estimation of the TX impairments at RXs and that $\Breve{\mathbf{y}}_{k}=\mathbf{0}_{M_k\times 1}$ (i$.$e$.$, yet no digital SI cancellation).
\begin{algorithm}[!t]
    \caption{Joint Digital Beamforming and SI Cancellation}
    \label{alg:the_alg}
    \begin{algorithmic}[1]
        \renewcommand{\algorithmicrequire}{\textbf{Input:}}
       \renewcommand{\algorithmicensure}{\textbf{Output:}}
        \REQUIRE $\mathbf{C}_k$, ${\rm P}_k$, ${\rm P}_m$, $\mu_1$, $\nu_1$, $\mu_2$, $\nu_3$, $\sigma_q^2$, and $\sigma_k^2$.
        \ENSURE $\mathbf{V}_k$, $\mathbf{G}_{1,k}$, $\mathbf{U}_k$, $\mathbf{V}_m$, $\mathbf{G}_{1,m}$, and $\mathbf{U}_q$.
        \STATE Obtain wireless channel estimates $\widehat{\mathbf{H}}_{k,k}, \widehat{\mathbf{H}}_{k,m}$, and $\widehat{\mathbf{H}}_{q,k}$ using pilot signals.
        \STATE Run \cite[Algorithm 1]{alexandropoulos2017joint} using $\widehat{\mathbf{H}}_{k,k}$ and $\widehat{\mathbf{H}}_{q,k}$, and perform column normalization to the result to obtain $\mathbf{V}_k$. Store $\mathbf{F}_k\in\mathbb{C}^{N_k\times \alpha}$ with $1\leq\alpha\leq\min\{M_q,N_k\}$ during this computation for step $5$.
        \STATE Set $[\mathbf{G}_{1,k}]_{(i,i)}=\sqrt{P_k/N_k}, \forall i=1,2,\dots,N_k$ and obtain $\mathbf{G}_{j,k}$ for $j=2,3,\ldots,6$ using \eqref{eq:small_gs}.
        \STATE Set $\mathbf{V}_m$ as the right-singular vectors of $\widehat{\mathbf{H}}_{k,m}$ and set $[\mathbf{G}_{1,m}]_{(i,i)}=\sqrt{P_m/N_m}, \forall i=1,2,\dots,N_m$.
        \STATE Set $\mathbf{U}_q$ as the left-singular vectors of the effective DL channel estimation matrix $\widehat{\mathbf{H}}_{q,k}\mathbf{F}_k$.
        \STATE Transmit the $L$ precoded pilot symbols $\mathbf{X}_{k}$ and $\mathbf{X}_{m}$.
        \STATE Measure the residual SI signal $\mathbf{J}_k$ in \eqref{eq:residual_signal} at the outputs of the RX RF chains of node $k$.
        \STATE Obtain estimate $\widehat{\Breve{\mathbf{H}}}_{k,k}$ using \eqref{eq:RQ_form} and $\boldsymbol{\widehat{\Omega}}=\widehat{\Breve{\mathbf{H}}}_{k,k}\mathbf{R}$, and set the digital SI cancellation signal as $\boldsymbol{\xi}_{k}$ in \eqref{eq:digital}.
    \end{algorithmic}
\end{algorithm}

Extending the design approach of \cite{alexandropoulos2017joint}, we focus on the estimated achievable FD rate defined as the sum of $\widehat{\mathcal{R}}_{\text{UL}}$ and $\widehat{\mathcal{R}}_{\text{DL}}$, and formulate the following general optimization problem for the joint design of the $N$-tap ($0\leq N< N_kM_k$) analog SI canceller and the digital TX/RX beamformers: 
\begin{align}\label{eq: optimization_eq}
        \nonumber\underset{\substack{\mathbf{C}_k,\mathbf{V}_k,\mathbf{U}_k,\\\mathbf{G}_{1,k},\mathbf{G}_{1,m},\\\mathbf{V}_m,\mathbf{U}_q}}{\text{max}} &\widehat{\mathcal{R}}_{\text{DL}}+\widehat{\mathcal{R}}_{\text{UL}}\\
        \text{\text{s}.\text{t}}\quad\quad &\mathbf{C}_k\,\, \text{as in}\,\,[7, {\rm eq. (C2)}],\nonumber\\
        &\mathbb{E}\{\|\mathbf{G}_{1,k}\mathbf{V}_k\mathbf{s}_k +\boldsymbol{\delta}_k\|^2\}\leq {\rm P}_k,\nonumber\\
        &\mathbb{E}\{\|\mathbf{G}_{1,m}\mathbf{V}_m\mathbf{s}_m\|^2\}\leq {\rm P}_m,\\
        &\mathbf{G}_{1,k},\mathbf{G}_{1,m}:\,\text{real}\,\,\text{diagonal}\,\,\text{matrices},\nonumber\\
        &\mathbf{V}_{k},\mathbf{U}_{k},\mathbf{V}_m,\mathbf{U}_{q}:\,\text{unit}\,\,\text{norm}\,\,\text{columns},\nonumber\\
        & \|(\widehat{\mathbf{H}}_{k,k}+\mathbf{C}_k )(\mathbf{G}_{1,k}\mathbf{V}_k\mathbf{s}_{k} + \boldsymbol{\delta}_k)\|^2<\lambda_{\rm A},\nonumber
\end{align}
where $\lambda_{\rm A}$ denotes the real-valued threshold residual power level after analog SI cancellation. This threshold depends on the dynamic range of the Analog-to-Digital Converter (ADC) and determines the saturation of the RF chains. A solution of \eqref{eq: optimization_eq} combined with a novel digital SI cancellation technique will be presented in the sequel.

\subsection{Proposed Digital SI Cancellation}
The solution of \eqref{eq: optimization_eq} determines the power of the SI signal after analog cancellation. As shown in Fig$.$~\ref{fig: transceiver}, we propose an additional digital cancellation block to boost SI cancellation in the digital domain. This block performs complementary to analog canceller ensuring SI suppression below RX noise floor.

To design digital SI cancellation, we use the baseband precoded samples at the TX of node $k$ to reconstruct the SI components, and then cancel them out at the output of each RX RF chain of the node. Supposing the $L$ samples $\mathbf{X}_{k} \triangleq\mathbf{V}_k\mathbf{S}_k\in\mathbb{C}^{N_k\times L}$ (and $\mathbf{X}_{m}\triangleq\mathbf{V}_{m}\mathbf{X}_{m}\in\mathbb{C}^{N_m\times L}$ with perfect synchronization), the residual SI signal $\mathbf{J}_{k}\in \mathbb{C}^{M_k \times L}$ after analog cancellation is derived using \eqref{eq: signal_DL} and \eqref{Eq:y_k_AC}-\eqref{Eq:y_k_SI} as 
\begin{equation}\label{eq:residual_signal}
        \mathbf{J}_k \triangleq\Breve{\mathbf{H}}_{k,k}\boldsymbol{\Psi}_k+\mathbf{H}_{k,m}\mathbf{G}_{1,m}\mathbf{X}_{m} + \mathbf{N}_{k},
\end{equation}
where $\Breve{\mathbf{H}}_{k,k}\in\mathbb{C}^{M_k \times 6N_k}$ is the residual SI matrix defined as $\Breve{\mathbf{H}}_{k,k}\triangleq \left(\mathbf{H}_{k,k}+\mathbf{C}_k \right)\mathbf{G}_{k}$ and 
$\boldsymbol{\Psi}_k\in\mathbb{C}^{6N_k\times L}$ contains the linear, image, and nonlinear SI signal components. The latter definitions have been obtained from \eqref{eq:augment_signal} for $L$ samples. 

The objective of the digital canceller is to estimate $\Breve{\mathbf{H}}_{k,k}$ and then subtract it at the ADC outputs of each RX RF chain. Making use of the SI samples in \eqref{eq:residual_signal} and the notation $\boldsymbol{\Theta}\in\mathbb{C}^{M_k \times 6N_k}$, the Least Squares (LS) estimation for $\Breve{\mathbf{H}}_{k,k}$ minimizing the power of the error matrix can be expressed as
\begin{equation}\label{eq:arg_min_res}
        \widehat{\Breve{\mathbf{H}}}_{k,k}\triangleq \argmin_{\boldsymbol{\Theta}}\|\mathbf{J}_k- \boldsymbol{\Theta}\mathbf{\Psi}_k\|^2.
\end{equation}  
This LS problem would have closed form solution if $\boldsymbol{\Psi}_k$ had independent rows. However, $\boldsymbol{\Psi}_k$'s rows are high order polynomials of linear and conjugate SI samples, as well as their interaction products. Therefore, its rows are correlated which results in high level of multi-collinearity. To tackle this problem, we propose orthogonalization of $\boldsymbol{\Psi}_k$ using RQ decomposition. We use traditional Gram-Schmidt orthogonalization to obtain $\boldsymbol{\Psi}_k=\mathbf{RQ}$, where $\mathbf{Q}\in \mathbb{C}^{L\times L}$ has orthogonal rows and $\mathbf{R}\in \mathbb{C}^{6N_k\times L}$ is a lower triangular matrix with ones on the main diagonal. Using the latter expression decomposition, \eqref{eq:arg_min_res} can be re-expressed as
\begin{equation}\label{eq:min_RQ}
    \begin{split}
        \widehat{\Breve{\mathbf{H}}}_{k,k}= \argmin_{\boldsymbol{\Theta}}\|\mathbf{J}_{k}- \boldsymbol{\Theta}\mathbf{RQ}\|^2.
    \end{split}
\end{equation}
Setting $\boldsymbol{\Omega}\triangleq\boldsymbol{\Theta}\mathbf{R}$ in \eqref{eq:min_RQ} results in a typical LS problem possessing the following closed form solution:
\begin{equation}\label{eq:RQ_form}
    \begin{split}
    \boldsymbol{\widehat{\Omega}}\triangleq&\argmin_{\boldsymbol{\Omega}} \|{\mathbf{J}_{k}-\boldsymbol{\Omega}\mathbf{Q}}\|^2
    = \mathbf{J}_{k} \mathbf{Q}^{\rm H} (\mathbf{QQ}^{\rm H})^{-1}.
    \end{split}
\end{equation}
Since $\boldsymbol{\widehat{\Omega}}=\widehat{\Breve{\mathbf{H}}}_{k,k}\mathbf{R}$, using back substitution yields $\widehat{\Breve{\mathbf{H}}}_{k,k}$.

At the FD MIMO node $k$, the digital SI cancellation block operates complementary to analog SI cancellation and digital TX/RX beamforming. It performs the estimation of $\Breve{\mathbf{H}}_{k,k}$ as described in \eqref{eq:residual_signal}--\eqref{eq:RQ_form} to derive the digital signal $\boldsymbol{\xi}_{k}$ in \eqref{Eq:Estimated_m} as
\begin{equation}\label{eq:digital}
    \begin{split}
        \boldsymbol{\xi}_{k}= -\widehat{\Breve{\mathbf{H}}}_{k,k}\boldsymbol{\psi}_k,
    \end{split}
\end{equation}
which is applied to the received signals after the ADCs. Our unified design of digital TX/RX beamforming with A/D SI cancellation includes the following two steps. We first find the $N$-tap analog SI canceller matrix $\mathbf{C}_k$ as described in \cite[Sec. III]{alexandropoulos2017joint}, and then obtain the digital TX/RX beamformers and digital SI cancellation signal as summarized in Algorithm \ref{alg:the_alg}. 

\section{Simulation Results and Discussion}\label{sec:Simulation}
In this section, we present performance evaluation comparisons among our proposed FD MIMO design and relevant ones intended for the communication scenario depicted in Fig. \ref{fig: transceiver}. 
\begin{table}[tbp]
    \caption{Simulation Parameters}
    \label{tab:1}
    \centering
    \scriptsize
    \begin{tabular}{|c|c|}
         \hline
         \!\!\!\textbf{Parameter}\!\!\! &\!\!\! \textbf{Value} \!\!\!\\
         \hline
         \!\!\!Signal Bandwidth\!\!\! &\!\!\! 1.4 MHz \!\!\!\\
         \!\!\!Constellation \!\!\!&\!\!\! 16-QAM\!\!\!\\
         \!\!\!No. of Subcarriers \!\!\!&\!\!\!  256\!\!\!\\
         \!\!\!No. of Data Subcarriers \!\!\!&\!\!\!  234\!\!\!\\
         \!\!\!Cyclic Prefix Length \!\!\!&\!\!\! 64\!\!\!\\
         \!\!\!TX Power \!\!\!&\!\!\! 20-40 dBm\!\!\!\\
         \hline
    \end{tabular}
    \!\!\!
    \begin{tabular}{|c|c|}
         \hline
         \!\!\!\textbf{Parameter} \!\!\!&\!\!\! \textbf{Value} \!\!\!\\
         \hline
         \!\!\!Node $k$ Noise Floor \!\!\!&\!\!\! $-$110.0 dBm\!\!\!\\
         \!\!\!Node $q$ Noise Floor \!\!\!&\!\!\! $-$90.0 dBm\!\!\!\\
         \!\!\!PA IIP3 \!\!\!&\!\!\! 15 dBm\!\!\!\\
         \!\!\!IRR \!\!\!&\!\!\! 30 dB\!\!\!\\
        \!\!\!No. of ADC Bits \!\!\!&\!\!\! 14\!\!\!\\
         \!\!\!PAPR \!\!\!&\!\!\! 10 dB\!\!\!\\
         \hline
    \end{tabular}
\end{table}

\begin{figure}[!tbp]
\centering
\includegraphics[width=.45\textwidth]{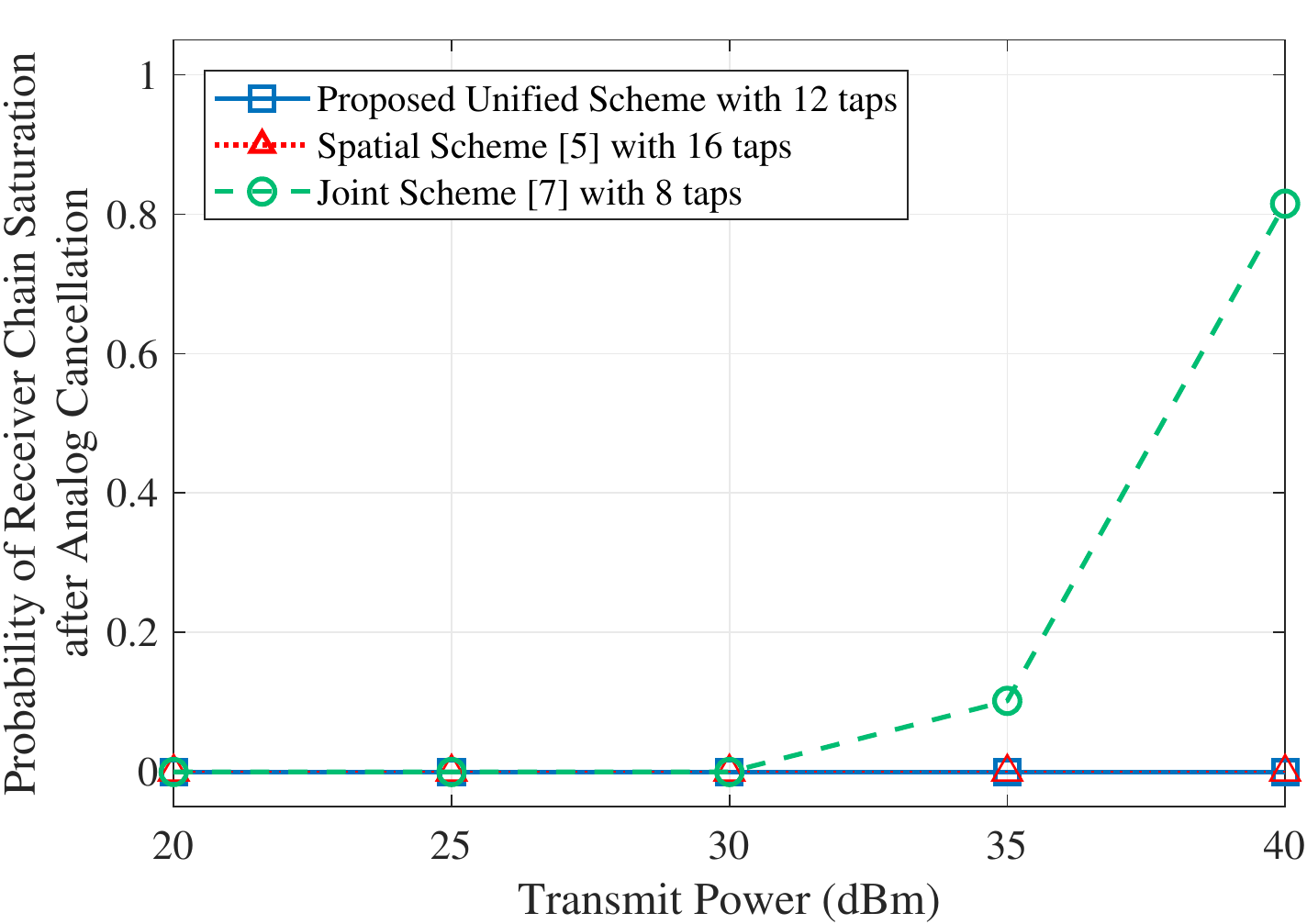}
\caption{Probability of saturation of any of the RX RF chains at the FD MIMO node $k$ after analog SI cancellation versus the TX power in dBm.}
\label{fig: prob_stauration}
\end{figure}
\subsection{Simulation Parameters}\label{subsec:sim_param}
We perform an extensive waveform simulation following the FD MIMO architecture illustrated in Fig$.$~\ref{fig: transceiver} and using baseband equivalent modeling of practical transceiver components including IQ imbalance and third-order PA nonlinearities \cite{ad9361}. We have considered $N_k=M_k=M_q=4$ for the TX and RX antennas at the FD node $k$ and the RX antennas of the half duplex node $q$. For the half duplex node $m$, we have assumed single symbol stream with $\mathbf{V}_m=N_m=d_m=1$. Both DL and UL channels are assumed as block Rayleigh fading channel with a pathloss of $110$ dB. The SI channel at the FD MIMO node $k$ is simulated as a Rician fading one with a $K$-factor of $35$ dB and pathloss of $40$ dB \cite{duarte2012experiment}. All involved wireless channels are estimated using pilot symbols before the symbol data communication\cite{sim2018self}. The generated waveforms were considered to be Orthogonal Frequency Division Multiplexing  (OFDM) signals with a bandwidth of $1.4$ MHz, which is a supported bandwidth for LTE. The additional parameters of the waveforms along with other system level parameters are shown in Table I. In order to incorporate TX impairments at the FD MIMO node $k$, the TX IRR was considered to be $30$ dB, which implies that the image signal power is $30$ dB lower than the linear SI signal \cite{korpi2014widely}. The quasi memoryless nonlinear PAs are assumed to have an IIP3 value of $15$ dBm \cite{ad9361}. Each ADC at the RX RF chains of node $k$ is considered to have $14$-bit resolution with an effective dynamic range of $62.24$ dB for a Peak-to-Average-Power-Ratio (PAPR) of $10$ dB \cite{AD3241}. Therefore, the residual SI power after analog cancellation has to be below $-47.76$ dBm to avoid saturation of each RX RF chain. Furthermore, non-ideal multi-tap analog canceller is considered with steps of $0.02$ dB for attenuation and $0.13^\circ$ for phase as in \cite{alexandropoulos2017joint,george2018journal}. We have used $1000$ Monte Carlo simulation runs ($1000$ independent set of channels) to calculate the performance of all considered designs. In each run, we have performed pilot-assisted estimation of all involved wireless channels and considered $200$ OFDM symbols transmitted from both nodes $k$ and $m$ to emulate a packet-based FD MIMO communication system.

\begin{figure}[!tbp]
\centering
\includegraphics[width=0.46\textwidth]{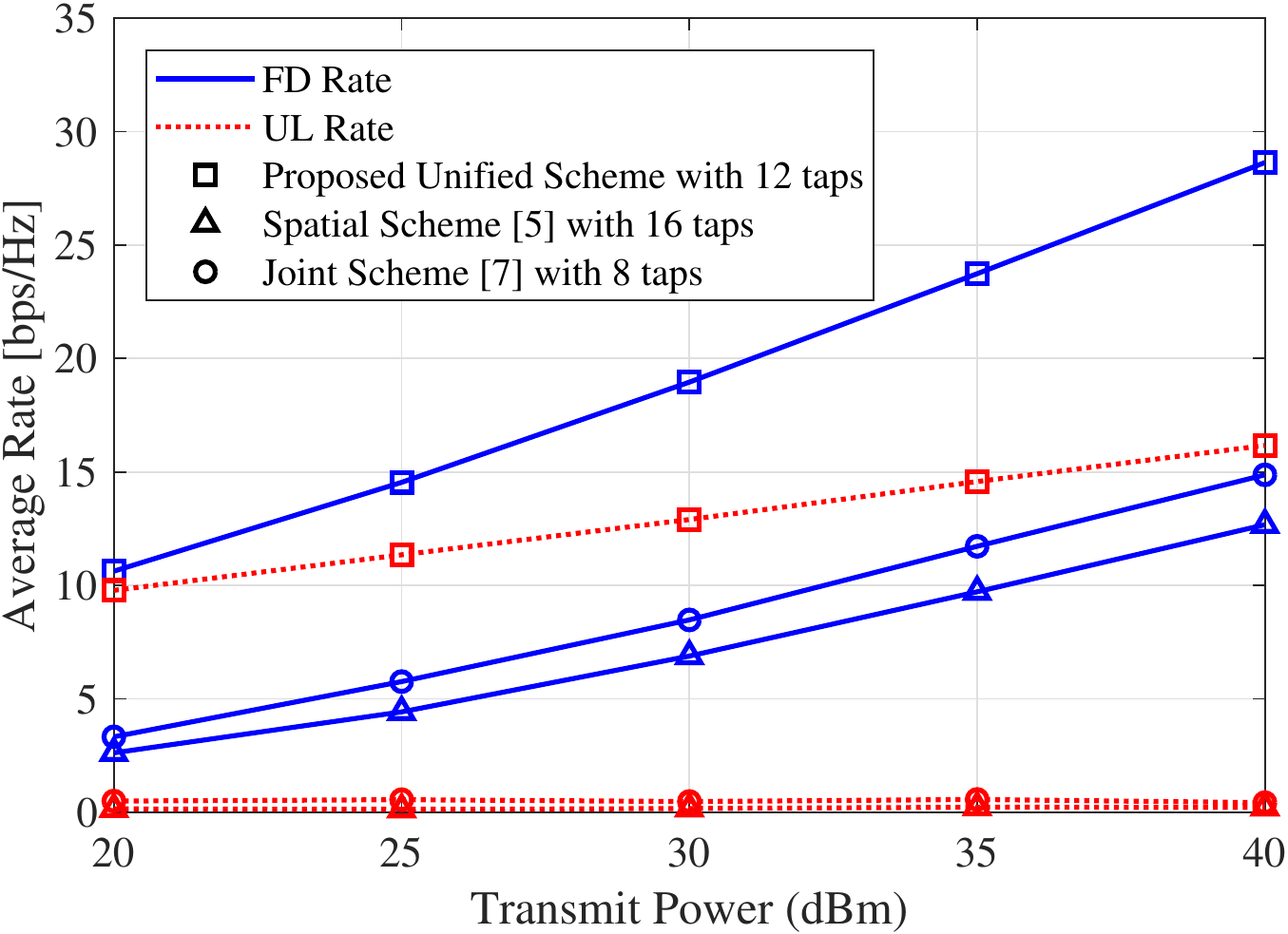}
\caption{Average UL and FD rates with respect to the TX power in dBm for the proposed and the state-of-the-art FD MIMO designs.}
\label{fig: full-duplex rate}
\end{figure}

\subsection{Compared FD MIMO Designs}\label{subsec:comp_design}
We compare the performance of our unified design of digital TX/RX beamforming and A/D SI cancellation with that of spatial suppression design presented in \cite{riihonen2011mitigation}. For the spatial suppression approach we have used a $16$-tap time domain analog SI canceller, which corresponds to the highest hardware complexity canceller for the considered number of transceiver antennas. In our unified approach we have incorporated both digital SI cancellation and an analog canceller with $12$ taps (i$.$e$.$, $25$\% less hardware for the canceller compared with the spatial suppression design), where the taps were placed in orderly column-by-column in $\mathbf{C}_k$. For the latter we use the term ``Proposed Unified Scheme with $12$ taps" and for the former the term ``Spatial Scheme [5] with $16$ taps" in the figures that follow. We have also considered the reduced complexity design of \cite{alexandropoulos2017joint} with only $8$ taps for the analog canceller (i$.$e$.$, $50$\% less hardware for the canceller compared with the spatial suppression design). This design performs joint analog SI cancellation with digital TX/RX beamforming, but does not include digital SI cancellation and transceiver RF chain impairments. We term this scheme as ``Joint Scheme [7] with $8$ taps" in the figures. 

We also present performance comparisons among our proposed digital SI cancellation technique and the widely linear digital cancellation proposed in \cite{korpi2014widely}, as well as the nonlinear one designed in \cite{bharadia2013full}. Both latter methods were extended to the MIMO case since they were initially proposed for SISO systems. In the results that follow we have computed the Bit Error Rate (BER) with $16$-ary Quadrature Amplitude Modulation (QAM) as well as the achievable UL and DL rates. These rates weres obtained from \eqref{eq:rates_eq} and \eqref{eq:covar_eq} after substituting the estimated channels with the actual ones.

\begin{figure}[!tbp]
\centering
\includegraphics[width=0.46\textwidth]{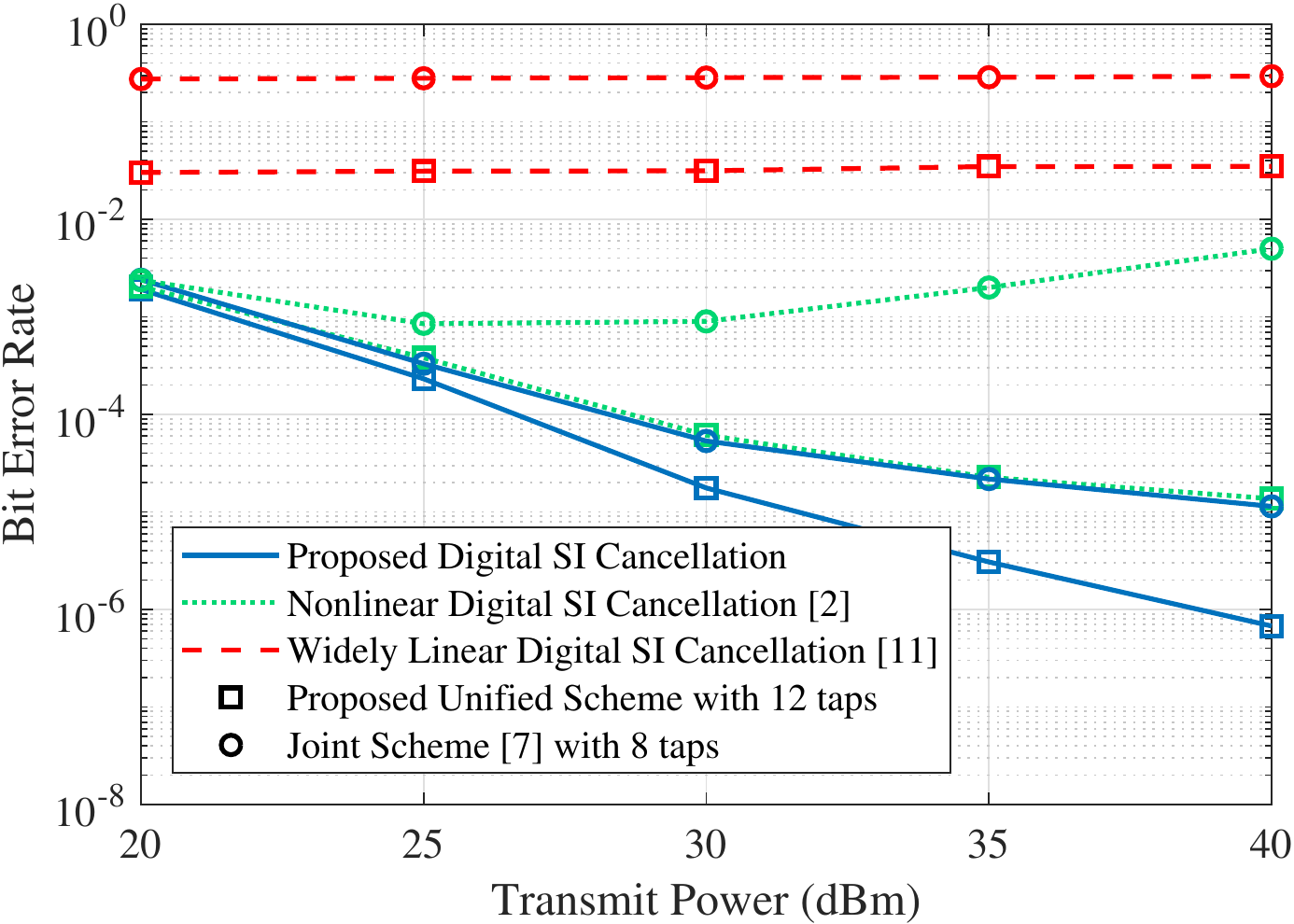}
\caption{Average BER with $16$-QAM at the FD MIMO node $k$ versus the TX power in dBm after cascading A/D SI cancellation designs.}
\label{fig: bit error rate}
\end{figure}
\subsection{SI Mitigation Capability, BER, and Achievable Rates}\label{subsec:rate_and_BER}

We start in Fig$.$~\ref{fig: prob_stauration} by comparing the SI mitigation capability of the three considered designs as a function of the TX power ${\rm P}_k$ in dBm. Recall that the residual SI after analog cancellation at the FD MIMO node $k$ should be below $-47.76$ dBm to avoid saturation of each RX RF chain. As shown, both the proposed scheme with $12$ taps and the spatial suppression approach with $16$ taps provide residual SI signal power below the threshold for the whole TX power range. However, this does not happen for the scheme of [7] with $8$ taps where digital SI cancellation is not considered. For TX power larger than $30$ dBm with this scheme, there exists non negligible probability that the residual SI power level is above the critical threshold $-47.76$ dBm.

The ergodic UL and FD rates as functions of the TX powers ${\rm P}_k$ and ${\rm P}_m$ in dBm are depicted in Fig$.$~\ref{fig: full-duplex rate}. It is evident that both schemes \cite{riihonen2011mitigation} and \cite{alexandropoulos2017joint} exhibit poor UL rates due to their inability to cancel the residual nonlinear and image SI signal components. However, our proposed unified scheme adequately suppresses all nonlinear SI components yielding increased UL performance. Additionally, the proposed scheme significantly outperforms both other in terms of achievable FD rate. In Fig$.$~\ref{fig: bit error rate}, we plot the BER with $16$-QAM versus ${\rm P}_k$ and ${\rm P}_m$ in dBm for both the proposed scheme and the scheme of \cite{alexandropoulos2017joint} with $12$- and $8$-tap analog SI cancellation, respectively. For both schemes we have considered three different digital SI cancellers: the proposed one, the widely linear \cite{korpi2014widely}, and the nonlinear  \cite{bharadia2013full} cancellers. Recall that \cite{alexandropoulos2017joint} was proposed without digital SI cancellation which we herein incorporate. It is shown that our proposed cancellation provides the best performance especially in high TX powers. More importantly, it is demonstrated that, even when the proposed digital SI cancellation is adopted at both schemes, the joint design of all blocks (digital TX/RX beamformers and A/D SI cancellation) is necessary for achieving the best performance.

\section{Conclusion and Future Work}
In this paper, we have presented a unified design of digital TX/RX beamforming with A/D SI cancellation for practical FD MIMO systems with TX IQ imbalances and PA nonlinearities. To suppress the residual linear SI signal along with its conjugate and nonlinear components, we proposed a novel digital SI cancellation technique that is jointly designed with the configuration of the multiple taps of the analog canceller and the digital TX/RX beamformers. Our selected performance evaluation comparisons with state-of-the-art designs demonstrated that our proposed approach is capable of achieving higher achievable rates and BER performance with reduced hardware complexity for analog SI cancellation. For future work, we intend to extend our unified design to wideband channels incorporating also practical RX impairments. 

\bibliographystyle{myIEEEtran}
\bibliography{mybib.bib}
\end{document}